%% file: main.tex
\documentclass[cameraready]{Interspeech}

\title{MeloDISinger: Melody-Aware \& Duration-Preserving Singing Voice Editing with Audio Infilling}

\author[affiliation={1}, orcid=0009-0003-4271-5987]{Yoonjeong}{Park}
\author[affiliation={2}, orcid=0009-0003-0996-3161]{Jaekwon}{Im}
\author[affiliation={1,2}, orcid=0000-0003-2664-2119]{Juhan}{Nam}

\address{
    $^1$ Graduate School of Artificial Intelligence, KAIST, South Korea \\
    $^2$ Graduate School of Culture Technology, KAIST, South Korea
}

\email{yoonjeong.park@kaist.ac.kr, jakeoneijk@kaist.ac.kr, juhan.nam@kaist.ac.kr}


\newcommand{\ours}{\textsc{MeloDISinger}}
\newcommand{\drp}{\textsc{MeloDRP}}

\usepackage{comment}
\usepackage{url}
\usepackage{multirow}
\usepackage{amsmath,cite,url}
\usepackage{graphicx}
\usepackage[table]{xcolor}
\usepackage{amssymb}
\usepackage{booktabs}
\usepackage{tabularx}

\begin{document}

\maketitle
\input{contents/0_abstract}
\input{contents/1_intro}
\input{contents/2_proposed}

\input{contents/3_exp}

\input{contents/4_result}
\input{contents/5_conclusion}
\input{contents/6_ack} 
\section{Generative AI Use Disclosure}
Generative AI tools were used solely to assist with English editing and polishing of the manuscript. 
All technical content, experimental design, implementation, results, and conclusions were produced and verified by the authors. 
No generative AI tool was used to generate new scientific claims, results, or to act as an author.

\bibliographystyle{IEEEtran}
\bibliography{mybib}

\end{document}

%% file: contents/0_abstract.tex
\begin{abstract}
\label{0_abstract}
Text-based singing voice editing (SVE) aims to revise sung lyrics while preserving the original melody, total duration, and non-edited regions. In this paper, we propose MeloDISinger, a flow-matching-based SVE model for melody-aware and duration-preserving editing. Its core module, MeloDRP, predicts fixed-budget
duration ratios, enabling explicit span-wise duration control. For melody-aware
duration allocation, MeloDRP fuses phonetic cues with pseudo-MIDI melodic context through cross-attention, while temporal-overlap supervision encourages soft phoneme--note correspondences. We further use a flow-matching mel decoder for audio infilling to synthesize edited regions while preserving surrounding context. In addition, we introduce a duration-aware edited-lyric generation pipeline using WhisperX and an LLM to construct feasible evaluation scenarios. Experiments demonstrate state-of-the-art performance in both objective and subjective evaluations.
\end{abstract}

%% file: contents/1_intro.tex
\section{Introduction}
\label{1_intro}
Singing voice plays a crucial role in music by conveying linguistic content and emotional expression \cite{sundberg2018singing}. 
As a musical component, it must remain aligned with the accompaniment in melody and rhythm. In music production, recorded vocals often require modifications, such as correcting mispronunciations, inserting missing words, or replacing specific phrases.
However, such modifications are typically costly and time-consuming, as they require professional studios, skilled vocalists, and specialized recording equipment. These challenges motivate the need for text-based singing voice editing.

Text-based singing voice editing (SVE) \cite{editsinger, vevo2, yingmusic, lei2024songcreator} aims to generate edited singing audio from original audio, original lyrics, and edited lyrics without manual waveform manipulation.
This task is similar to speech editing, which modifies speech segments to match a revised transcript\cite{alexos2024attentionstitchattentionsolvesspeech, peng2024voicecraft, voicebox}. In both tasks, only the audio segments corresponding to the revised text should be regenerated, while the remaining regions should be preserved with seamless transitions at the edit boundaries.
Unlike speech editing, however, SVE is more challenging due to additional music-specific constraints. Edited segments must remain consistent with the original melody and rhythm, and the total duration of the edited audio must be strictly preserved to maintain synchronization with the accompaniment.

Previous SVE methods fall into two categories for duration and pitch modeling: implicit and explicit. In both paradigms, most existing methods\cite{editsinger, vevo2, lei2024songcreator} do not explicitly guarantee strict preservation of the total duration, which can result in temporal misalignment with the accompaniment. Implicit approaches\cite{vevo2, yingmusic, lei2024songcreator} learn text-audio alignments from large-scale datasets without requiring phoneme-level annotations, and synthesize edited singing voices conditioned on prosody tokens or melody features. However, such conditioning does not impose hard constraints on duration or melody, making it difficult to strictly preserve the total duration and follow the original rhythmic and melodic structure. Since these methods often regenerate the entire sequence, they may also alter timing or melody in non-edited regions. In contrast, EditSinger~\cite{editsinger} explicitly models duration and pitch using variance adaptors~\cite{fastspeech2}, but its duration modeling lacks melodic context which can produce speech-like timing inconsistent with the original rhythm. Furthermore, it reuses original phoneme durations for replacing phonemes, restricting replacement to cases with the same number of phonemes and potentially causing unnatural pronunciation. In addition, EditSinger does not explicitly address seamless integration at edit boundaries. 

In this paper, we propose \ours{}
(\textbf{Melo}dy-aware \textbf{D}uration-ratio \textbf{I}nfilling for singing
voice editing), a flow-matching-based SVE model for melody-aware and
duration-preserving editing. Its key component, \drp{} (\textbf{Melo}dy-aware
\textbf{D}uration \textbf{R}atio \textbf{P}redictor), predicts fixed-budget
duration ratios rather than absolute durations, enabling explicit span-wise
duration control while preserving the original duration of each edit span. For
melody-aware duration allocation, \drp{} fuses phonetic cues with pseudo-MIDI
melodic context through cross-attention, with temporal-overlap supervision
encouraging soft phoneme--note correspondences. We also employ a flow-matching-based mel decoder in an
audio-infilling manner to synthesize edited regions while preserving non-edited
regions. In addition, we introduce an automatic duration-aware edited-lyric
generation pipeline using WhisperX-based alignment and an LLM to construct
temporally feasible evaluation scenarios. Experimental results show that \ours{}
achieves state-of-the-art performance in both objective and subjective
evaluations.

%% file: contents/2_proposed.tex
\begin{figure*}[t]
  \centering
  \includegraphics[width=\linewidth, height=0.3\textheight]{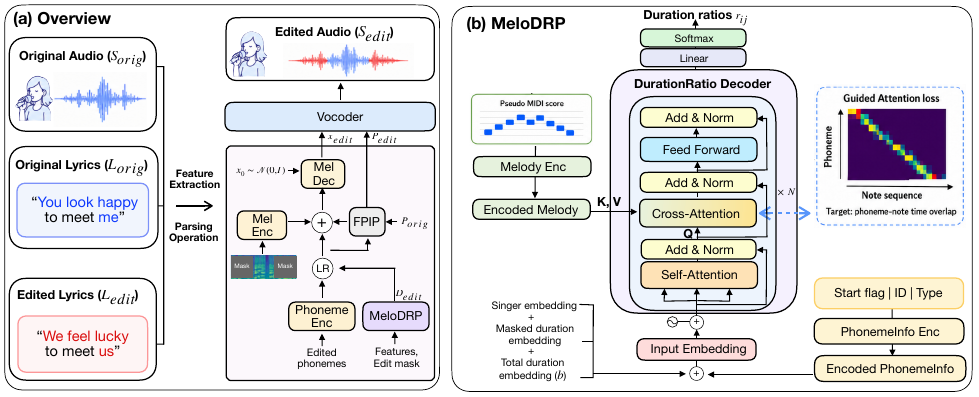}
  \caption{Overview of \ours{}: (a) overall text-based SVE pipeline and
(b) \drp{} architecture for melody-aware duration-ratio prediction.}
  \label{fig:overall}
\end{figure*}

\section{Method}
\label{3_method}

\subsection{Overview}
As shown in Fig.~\ref{fig:overall}, \ours{} follows the three-step pipeline of
EditSinger~\cite{editsinger}: feature extraction, parsing operation, and modeling.
Let $S_{\mathrm{orig}}$, $S_{\mathrm{edit}}$, $L_{\mathrm{orig}}$, and $L_{\mathrm{edit}}$ denote the original audio, edited audio, original lyrics, and edited lyrics, respectively.

\noindent \textbf{Feature Extraction.}
From $S_{\mathrm{orig}}$, we extract acoustic features: mel-spectrogram, speaker embedding, frame-level F0 with voiced/unvoiced flags and a pseudo score derived from F0. From $S_{\mathrm{orig}}$ and $L_{\mathrm{orig}}$, we obtain original phoneme durations using the Montreal Forced Aligner~\cite{mfa}. We convert $L_{\mathrm{edit}}$ into phonemes using g2p-en and derive two phoneme-level linguistic features: start flags (2/1/0 for word-initial, syllable-initial but non-word-initial, and  others) and coarse phoneme types based on manner of articulation and vowel stress.

\noindent \textbf{Parsing Operation.}
To localize edit regions, we compare $L_{\mathrm{orig}}$ and $L_{\mathrm{edit}}$ to obtain edit types and a phoneme-level edit mask. 

\noindent \textbf{Modeling.}
Given the extracted features and edit mask, \ours{} synthesizes $S_{\mathrm{edit}}$ using an acoustic model--vocoder framework, widely used in TTS~\cite{tacotron, Eskimez2024E2TE, chen-etal-2024-f5tts} and SVS~\cite{diffsinger,Zhang_2024}. For explicit control over timing and pitch in the edited regions, the acoustic model predicts edited phoneme duration ($D_{edit}$) with \drp{} and the edited-region F0 contour ($P_{edit}$) with FPIP, adopted from EditSinger~\cite{editsinger}.

\subsection{\drp{}}
\label{sec:melodrp}
Conventional duration predictors\cite{editsinger, visinger,xiaoicesing} estimate absolute phoneme durations without inherently enforcing a prescribed total duration.
In SVE, however, each edited region must fit the original time span to synchronize with the accompaniment. We therefore propose \textbf{\drp{}}, a \textbf{Melo}dy-aware \textbf{D}uration \textbf{R}atio \textbf{P}redictor that reformulates duration modeling as span-wise duration reallocation under a fixed total-duration budget.

Given $N$ disjoint edit spans, let $T_i$ be the duration budget of span $i$, and let $E_i$ be the number of target phonemes in the span. \drp{} predicts duration ratios $\{r_{ij}\}_{j=1}^{E_i}$ and recovers phoneme durations by
\begingroup
\setlength{\abovedisplayskip}{3pt plus 1pt minus 1pt}
\setlength{\belowdisplayskip}{3pt plus 1pt minus 1pt}
\setlength{\abovedisplayshortskip}{1pt plus 1pt minus 1pt}
\setlength{\belowdisplayshortskip}{3pt plus 1pt minus 1pt}
\begin{equation}
    \sum\nolimits_{j=1}^{E_i} r_{ij} = 1, \quad
    \hat{d}_{ij} = T_i r_{ij},
\label{eq:duration_ratio}
\end{equation}
\endgroup
where $\hat{d}_{ij}$ denotes the predicted duration of the $j$-th phoneme in span $i$. This preserves each span duration by construction, since $\sum_{j=1}^{E_i}\hat{d}_{ij}=T_i$.
The budget $T_i$ is operation-dependent: (i) replacement reallocates the replaced-span duration to the replacing phonemes, (ii) insertion reallocates neighboring-word duration to neighboring and inserted phonemes, and (iii) deletion assigns the deleted-span duration to a silence phoneme. We provide $T_i$ to \drp{} as a phoneme-level budget sequence $\mathbf{b}$, with $b_p=T_i$ inside span $i$ and $b_p=0$ elsewhere.

Since singing phoneme durations are strongly tied to rhythm and melody{\cite{visinger, wang2022singingtacotron}, relying only on phoneme representations and surrounding durations can yield speech-like timing inconsistent with the original melody. To enable melody-aware duration prediction, we enrich the edited phoneme sequence with phoneme-type embeddings and word/syllable-initial indicators to provide articulatory and linguistic-boundary cues.
In \drp{}, the resulting phoneme-side
representation is fused with encoded pseudo-MIDI representations through cross-attention. We extract pseudo-MIDI from the original singing audio rather than score annotations, since performed singing often deviates from the score in both pitch and timing~\cite{hifisinger}. The fused representations are projected to duration logits and normalized by a span-wise softmax, enforcing $\sum_j r_{ij}=1$ within each edit span. Rather than imposing hard one-to-one phoneme--note alignment, \drp{} learns soft
phoneme--note correspondences from phonetic cues, onset cues, note timing,
duration-ratio losses, and a guided-attention loss based on phoneme--note
temporal overlap.

The \drp{} loss is defined as
\begingroup
\setlength{\abovedisplayskip}{3pt plus 1pt minus 1pt}
\setlength{\belowdisplayskip}{3pt plus 1pt minus 1pt}
\setlength{\abovedisplayshortskip}{1pt plus 1pt minus 1pt}
\setlength{\belowdisplayshortskip}{3pt plus 1pt minus 1pt}
\begin{equation}
\mathcal{L}_{\mathrm{MeloDRP}}
=
\sum\nolimits_{q\in\mathcal{Q}} \lambda_q \mathcal{L}_q,
\quad
\mathcal{Q}=\{\mathrm{ph},\mathrm{wd},\mathrm{pen},\mathrm{ga}\}.
\end{equation}
\endgroup
where the $\lambda$ terms are weighting hyperparameters. For span $i$, let $D_i^{*}=\sum_{k=1}^{E_i}d^{*}_{ik}$. The target ratio is
$r^{*}_{ij}=d^{*}_{ij}/D_i^{*}$. 
$\mathcal{L}_{\mathrm{ph}}$ is a KL divergence
$\mathrm{KL}(\mathbf{r}^{*}_i \Vert \mathbf{r}_i)$, and
$\mathcal{L}_{\mathrm{wd}}$ applies an L1 loss after aggregating phoneme ratios
within each word. $\mathcal{L}_{\mathrm{pen}}$ penalizes edit-span phonemes whose
predicted duration $\hat{d}_{ij}=T_i r_{ij}$ is shorter than a predefined threshold $d_{\min}$.
Unlike diagonal
guided-attention priors~\cite{GA_1,GA_2}, $\mathcal{L}_{\mathrm{ga}}$ uses an L1
loss to match cross-attention to a binary
phoneme--note temporal-overlap mask constructed
from ground-truth phoneme durations and pseudo-MIDI note durations.

\subsection{Flow-Matching-Based Mel Decoder with Infilling}
\label{sec:mel_decoder}
We adopt an audio infilling{~\cite{Eskimez2024E2TE, voicebox, wu2024laugh} scheme to generate the edited mel-spectrogram, enabling seamless transitions at edit boundaries while preserving the non-edited context.
The decoder is a non-autoregressive conditional flow-matching model conditioned on the sum of frame-level phoneme, pitch, speaker, and context-mel embeddings.
During training, we sample random edit masks and train the decoder only on the
masked regions. Let $x_0\sim\mathcal{N}(0,I)$, $x_1=x_{\mathrm{orig}}$, and
$u=x_1-x_0$. We sample $t\sim\mathcal{U}(0,1)$ and use the linear probability
path $x_t=(1-t)x_0+t x_1$.  We jointly optimize the mel decoder, mel encoder, and phoneme encoder with the
conditional flow-matching objective{~\cite{lipman2022flow}}:
\begingroup
\setlength{\abovedisplayskip}{3pt plus 1pt minus 1pt}
\setlength{\belowdisplayskip}{3pt plus 1pt minus 1pt}
\setlength{\abovedisplayshortskip}{1pt plus 1pt minus 1pt}
\setlength{\belowdisplayshortskip}{3pt plus 1pt minus 1pt}
\begin{equation}
\begin{aligned}
\mathcal{L}_{\mathrm{CFM}}
=
\mathbb{E}_{x_1,t,x_0}
\Bigl[
\lVert
\bigl(
v_\theta(x_t,t,c)
-
u
\bigr)
\odot m_{\mathrm{edit}}
\rVert_2^2
\Bigr].
\end{aligned}
\end{equation}
\endgroup
where $m_{\mathrm{edit}}$ is the frame-level edit mask.
At inference, $c$ is constructed from the edited phoneme sequence, durations
predicted by \drp{}, pitch predicted by FPIP, speaker embedding, and the original
mel context. Starting from Gaussian noise, we solve the learned ODE to obtain
$\hat{x}_{\mathrm{gen}}$, and merge its edited frames with the original
mel-spectrogram:
\begin{equation}
x_{\mathrm{edit}}
=
m_{\mathrm{edit}}\odot \hat{x}_{\mathrm{gen}}
+
(1-m_{\mathrm{edit}})\odot x_{\mathrm{orig}}.
\end{equation}
This preserves the non-edited regions exactly while allowing the edited region
to be generated with surrounding context.
\subsection{Evaluation Set Generation Pipeline}
\label{method:pipeline}
We propose a duration-aware edited-lyric generation pipeline for evaluation. Prior SVE studies \cite{vevo2,yingmusic} typically generate edited lyrics by applying LLM-based rewriting to the original lyrics only, following speech-editing-style evaluation protocols. However, lyric-only rewriting may produce temporally infeasible edits in singing, since the edited lyrics must fit within the available singing duration.
To address this, we first estimate word-level onset and offset times from the original singing audio $S_{\mathrm{orig}}$ using WhisperX~\cite{whisperx}. We then convert each timing slot into a syllable capacity, which specifies the
maximum number of syllables that can be sung within the slot:
\begingroup
\setlength{\abovedisplayskip}{3pt plus 1pt minus 1pt}
\setlength{\belowdisplayskip}{3pt plus 1pt minus 1pt}
\setlength{\abovedisplayshortskip}{1pt plus 1pt minus 1pt}
\setlength{\belowdisplayshortskip}{3pt plus 1pt minus 1pt}
\begin{equation}
C = \left\lfloor \alpha \Delta t / \tau_{\min} \right\rfloor,
\end{equation}
\endgroup
where $\Delta t$ is the available duration, $\tau_{\min}$ is the minimum stable
per-syllable singing duration, and $\alpha$ is a safety margin for alignment
noise. We set $\tau_{\min}=0.3\,\mathrm{s}$ and $\alpha=0.8$. For replacement, $\Delta t$ is defined from the word or phrase being replaced. For insertion, $\Delta t$ is defined from the local span around the insertion point, and the syllables of neighboring anchor words are subtracted from the capacity.
Finally, we provide the original lyrics, edit instructions, and syllable-capacity metadata to an LLM (Gemini-2.5-flash) to generate edited lyrics that satisfy both the target edit scenario and the temporal constraints. Each candidate is automatically verified, and invalid candidates are regenerated.

%% file: contents/3_exp.tex
\input{tables/main_results_table}
\section{Experiments}
\subsection{Experimental Setup}
\noindent \textbf{Dataset and preprocessing.}
We conduct experiments on GTSinger-En~\cite{gtsinger}, which contains 13 hours of
English singing voices from three singers. Each audio sample is segmented into
chunks of up to 11.6 seconds, corresponding to 1000 mel frames, while preserving
word boundaries using phoneme duration annotations. Following the PC-NSF
HiFi-GAN\footnote{\url{https://github.com/openvpi/vocoders/releases}} vocoder setting, we use a sampling rate of 44.1 kHz, window size of
2048, hop size of 512, and 128 mel bins. We extract speaker embeddings and F0
using Resemblyzer\footnote{\url{https://github.com/resemble-ai/Resemblyzer}} and
Parselmouth\footnote{\url{https://github.com/YannickJadoul/Parselmouth}},
respectively. Frames with F0 $<3$ are treated as unvoiced following
\cite{hifisinger}, and pseudo-MIDI scores are derived from F0 by MIDI
quantization, note segmentation and note-level post-processing. 
\noindent \textbf{Model configuration.}
The phoneme and melody encoders are 4-layer Transformer encoders, and the
duration-ratio decoder is a 6-layer Transformer decoder. All Transformer modules
use hidden size 256 and 2 attention heads. The mel encoder is an MLP, and following \cite{diffsinger}, the mel
decoder uses a non-causal WaveNet~\cite{wavenet} with 20 residual layers and 256 channels. 

\noindent \textbf{Training and inference.}
We train the model using Adam with $\beta_1{=}0.9$, $\beta_2{=}0.999$, learning
rate $1{\times}10^{-4}$, and batch size 16. The learning rate is decayed by 0.5
at 10k, 20k, and 30k steps using MultiStepLR.
During reconstruction-based
training, we sample phoneme-level edit masks with ratio
$r\sim\mathcal{U}(0.3,0.7)$, allowing both continuous and discontinuous masked
spans. We apply phoneme identity/type embedding dropout with probability 0.3. At
inference, the flow-matching decoder is sampled with 100 Euler steps.

\subsection{Baselines and Evaluation Metrics}
\noindent \textbf{Evaluation scenarios.} Using the pipeline in Sec.~\ref{method:pipeline}, we construct six edited-lyric
sets for each original lyric: insertion (Ins), deletion (Del), mixed edits (Mix), and three
replacement settings. The replacement settings are
phoneme-matched (Rep-P), where the phoneme count is unchanged;
syllable-matched (Rep-S), where the syllable count is unchanged but the phoneme
count differs; and syllable-mismatched (Rep-SM), where both phoneme and
syllable structures differ. For single-operation settings, we edit at least three word positions when the
lyric contains four or more words.

\noindent \textbf{Baselines.}
We evaluate on 60 clips sampled from eight unseen GTSinger-En songs covering all
six singing techniques. We compare \ours{} with EditSinger~\cite{editsinger} and
Vevo2~\cite{vevo2}. Since the official implementation of EditSinger is not
publicly available, we reproduce it based on the paper.

\noindent \textbf{Objective evaluation.}
We evaluate intelligibility using WER and CER computed by
Whisper-large-v3~\cite{whisper}, duration preservation using Duration
Consistency (DC)~\cite{vevo2} and Duration Difference (DDUR), and melody
following using F0 Pearson Correlation (FPC) on voiced regions
\cite{vevo2,yingmusic,fpc_1}. Since baselines may change the edited-region
duration, we report two FPC variants: FPC-Cut, which truncates F0 sequences to
the shorter length, and FPC-DTW, which aligns them using DTW.

\noindent \textbf{Subjective evaluation.}
We conduct MOS tests with 22 listeners. Given the original audio, original lyrics,
and edited lyrics, listeners rate each generated sample on Lyric Following, Melody Following, and Naturalness using a 1--5 scale with 0.5-point
increments.

%% file: tables/main_results_table.tex
\begin{table}[t]
\caption{Objective results. Best values are bold.}
\label{tab:main_results}
\centering
\footnotesize
\renewcommand{\arraystretch}{0.95}
\setlength{\tabcolsep}{1.2pt}
\begin{tabular}{@{}llcc@{\hspace{0.80em}}cc@{\hspace{0.80em}}cc@{}}
\toprule
Set & Model
& \multicolumn{2}{c}{Intell.}
& \multicolumn{2}{c}{Dur.}
& \multicolumn{2}{c}{FPC} \\
\cmidrule(lr){3-4} \cmidrule(lr){5-6} \cmidrule(lr){7-8}
& & WER$\downarrow$ & CER$\downarrow$
& DDUR$\downarrow$ & DC$\uparrow$
& Cut$\uparrow$ & DTW$\uparrow$ \\
\midrule

\multirow{3}{*}{Rep-P}
& EditSinger & 38.80 & 27.26 & \textbf{0.00} & \textbf{99.93} & 74.13 & \textbf{71.54} \\
& Vevo2 & 51.45 & 39.32 & 0.59 & 87.72 & 28.43 & 51.17 \\
& MeloDISinger & \textbf{31.33} & \textbf{20.98} & \textbf{0.00} & \textbf{99.93} & \textbf{75.93} & 66.03 \\
\midrule

\multirow{2}{*}{Rep-S}
& Vevo2 & 42.29 & 30.90 & 1.04 & 82.46 & 14.12 & 46.23 \\
& MeloDISinger & \textbf{21.88} & \textbf{15.26} & \textbf{0.00} & \textbf{99.93} & \textbf{69.90} & \textbf{63.12} \\
\midrule

\multirow{2}{*}{Rep-SM}
& Vevo2 & 40.89 & 32.95 & 1.30 & 77.71 & 10.79 & 47.06 \\
& MeloDISinger & \textbf{28.74} & \textbf{21.16} & \textbf{0.00} & \textbf{99.93} & \textbf{68.87} & \textbf{63.17} \\
\midrule

\multirow{3}{*}{Ins}
& EditSinger & 19.67 & 12.16 & 0.55 & 89.97 & 61.76 & 70.14 \\
& Vevo2 & 31.43 & 21.73 & 1.43 & 73.51 & 1.72 & 50.28 \\
& MeloDISinger & \textbf{18.57} & \textbf{11.62} & \textbf{0.00} & \textbf{99.93} & \textbf{77.71} & \textbf{71.14} \\
\midrule

\multirow{3}{*}{Del}
& EditSinger & 27.01 & 16.23 & 0.67 & 87.64 & 23.37 & 75.59 \\
& Vevo2 & 68.72 & 53.90 & 1.92 & 70.56 & 0.18 & 37.64 \\
& MeloDISinger & \textbf{24.88} & \textbf{15.74} & \textbf{0.00} & \textbf{99.93} & \textbf{94.74} & \textbf{80.53} \\
\midrule

\multirow{2}{*}{Mix}
& Vevo2 & 50.93 & 38.38 & 0.79 & 84.49 & 20.70 & 38.50 \\
& MeloDISinger & \textbf{39.38} & \textbf{27.63} & \textbf{0.00} & \textbf{99.93} & \textbf{67.65} & \textbf{48.67} \\
\bottomrule
\end{tabular}
\vspace{-1em}
\end{table}

%% file: contents/4_result.tex
\input{tables/subj_eval_table}
\input{tables/ablations_results_table}
\section{Results}
\subsection{Objective Evaluation}
Table~\ref{tab:main_results} reports the objective results across all edit
scenarios. All metrics are reported in \% except DDUR, which is reported in
seconds; DDUR values shown as 0.00 correspond to a residual deviation of about
0.004\,s due to the STFT hop size. Overall, \ours{} achieves the best performance
in most metrics and scenarios, while strictly preserving the total duration. In contrast,
EditSinger and Vevo2 show non-zero DDUR, which can lead to temporal misalignment
with the accompaniment.
Compared with EditSinger, \ours{} achieves better intelligibility in most
replacement settings, even though it is not restricted to phoneme-matched edits.
This suggests that reusing original phoneme durations for edited phonemes can
distort articulation when the phonetic composition changes, whereas \drp{}
reallocates durations according to both phonetic and melodic context. Compared
with Vevo2, \ours{} substantially reduces WER/CER and improves FPC, indicating
more accurate lyric following and stronger melody preservation. 

\subsection{Subjective Evaluation}
Table~\ref{tab:subjective_mos} shows that \ours{} achieves the highest MOS across
all criteria and scenarios. The gains are particularly large in Rep-SM and Mix,
where duration reallocation must handle changed phonetic and syllabic structures
while preserving the original melody. Compared with EditSinger, \ours{} improves
perceptual quality even in Rep-P, suggesting that reusing original phoneme
durations is insufficient for natural SVE. EditSinger's lower Melody Following
score in Ins is consistent with speech-like timing in inserted regions. Compared
with Vevo2, \ours{} yields more reliable lyric rendering, melody following, and
naturalness.

\subsection{Ablations}

\begin{figure}[t]
  \centering  \includegraphics[width=\linewidth]{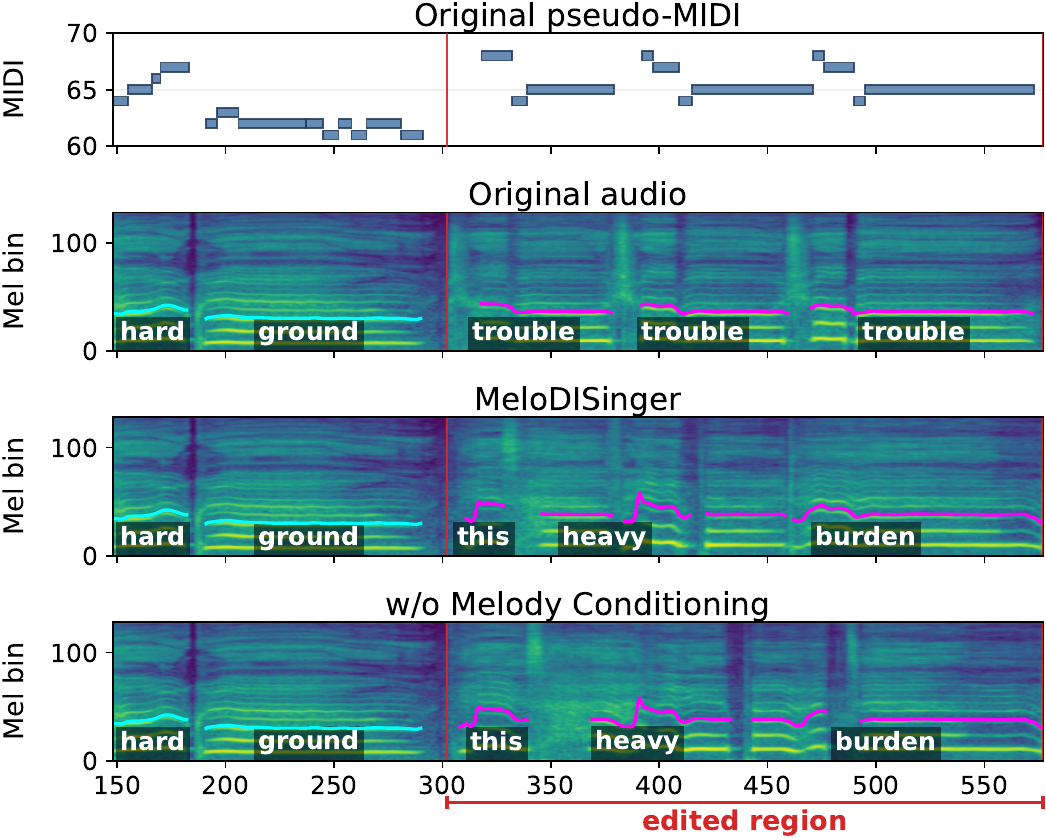}
  \vspace{-1.5em}
  \caption{Qualitative effect of melody conditioning.}
  \label{fig:melody_ablation}
  \vspace{-1.5em}
\end{figure}
Table~\ref{tab:ablation_results} shows the effect of each \drp{} component. Full
denotes \ours{}, and -Mel, -GA, -Phn, and -Dur remove melody conditioning,
guided-attention loss, phoneme information, and total-duration conditioning,
respectively. Removing -Dur causes the largest and most consistent degradation,
confirming that duration-ratio prediction must account for the available edit-span
budget. Removing -Mel also degrades Rep-S, Rep-SM, Ins, and Mix, where duration
allocation must follow the original rhythmic and melodic structure.
Fig.~\ref{fig:melody_ablation} qualitatively shows that melody conditioning helps
the edited-region timing and F0 contour follow the time-overlapping pseudo-MIDI
note structure. Removing -Phn mainly harms Ins, where inserted phonemes require
articulatory cues. 
The guided-attention loss has a smaller and more scenario-dependent effect in WER/CER, but without it, the predicted duration ratios become less sensitive to melodic context.

%% file: tables/subj_eval_table.tex
\begin{table}[t]
\caption{Subjective MOS with 95\% confidence intervals. Best values are bold.}
\label{tab:subjective_mos}
\centering
\footnotesize
\setlength{\tabcolsep}{3.0pt}
\renewcommand{\arraystretch}{0.95}
\begin{tabular}{@{}llccc@{}}
\toprule
Set & Model
& \multicolumn{3}{c}{MOS} \\
\cmidrule(lr){3-5}
& & Lyric Fol. & Melody Fol. & Naturalness \\
\midrule

\multirow{3}{*}{Rep-P}
& EditSinger & 2.64$\pm$0.21 & 3.24$\pm$0.21 & 2.40$\pm$0.17 \\
& Vevo2 & 2.45$\pm$0.24 & 2.86$\pm$0.20 & 2.55$\pm$0.18 \\
& MeloDISinger & \textbf{3.66$\pm$0.21} & \textbf{3.35$\pm$0.21} & \textbf{3.10$\pm$0.18} \\
\midrule

\multirow{2}{*}{Rep-S}
& Vevo2 & 3.80$\pm$0.25 & 2.53$\pm$0.29 & 3.08$\pm$0.19 \\
& MeloDISinger & \textbf{4.26$\pm$0.18} & \textbf{3.83$\pm$0.20} & \textbf{3.85$\pm$0.16} \\
\midrule

\multirow{2}{*}{Rep-SM}
& Vevo2 & 2.92$\pm$0.33 & 3.08$\pm$0.27 & 2.86$\pm$0.23 \\
& MeloDISinger & \textbf{4.05$\pm$0.23} & \textbf{3.99$\pm$0.20} & \textbf{3.65$\pm$0.19} \\
\midrule

\multirow{3}{*}{Ins}
& EditSinger & \textbf{3.95$\pm$0.23} & 2.92$\pm$0.26 & 3.01$\pm$0.24 \\
& Vevo2 & 2.14$\pm$0.26 & 2.83$\pm$0.29 & 2.11$\pm$0.22 \\
& MeloDISinger & \textbf{3.95$\pm$0.19} & \textbf{3.55$\pm$0.22} & \textbf{3.27$\pm$0.19} \\
\midrule

\multirow{3}{*}{Del}
& EditSinger & 4.03$\pm$0.22 & 3.75$\pm$0.24 & 3.45$\pm$0.21 \\
& Vevo2 & 2.36$\pm$0.29 & 1.58$\pm$0.17 & 2.69$\pm$0.28 \\
& MeloDISinger & \textbf{4.21$\pm$0.25} & \textbf{4.05$\pm$0.24} & \textbf{3.87$\pm$0.22} \\
\midrule

\multirow{2}{*}{Mix}
& Vevo2 & 2.13$\pm$0.24 & 2.31$\pm$0.29 & 2.39$\pm$0.27 \\
& MeloDISinger & \textbf{4.12$\pm$0.22} & \textbf{3.64$\pm$0.22} & \textbf{3.48$\pm$0.17} \\
\bottomrule
\end{tabular}
\end{table}

%% file: tables/ablations_results_table.tex
\begin{table}[!t]
\caption{Ablation results across edit scenarios (WER/CER). Best values are bold.}
\label{tab:ablation_results}
\centering
\footnotesize
\setlength{\tabcolsep}{0pt}
\renewcommand{\arraystretch}{0.95}
\newcommand{\best}[1]{\textbf{#1}}
\begin{tabular}{
@{}l
@{\hspace{1em}}r@{\,/\,}l
@{\hspace{1em}}r@{\,/\,}l
@{\hspace{1em}}r@{\,/\,}l
@{\hspace{1em}}r@{\,/\,}l
@{\hspace{1em}}r@{\,/\,}l
@{}}
\toprule
Config.
& \multicolumn{2}{c}{Rep-P}
& \multicolumn{2}{c}{Rep-S}
& \multicolumn{2}{c}{Rep-SM}
& \multicolumn{2}{c}{Ins}
& \multicolumn{2}{c}{Mix} \\
\midrule
Full
& \best{31.3} & 21.0
& \best{21.9} & \best{15.3}
& \best{28.7} & \best{21.2}
& 18.6 & \best{11.6}
& \best{39.4} & 27.6 \\
\addlinespace[0.2em]

-Mel
& \best{31.3} & \best{20.7}
& 23.1 & 16.4
& 30.6 & 22.2
& 20.8 & 13.8
& 43.7 & 29.5 \\

-GA
& 33.2 & 22.4
& 22.3 & 16.0
& 32.8 & 23.6
& \best{17.8} & \best{11.6}
& 39.8 & \best{26.3} \\

-Phn
& 32.6 & 22.2
& 22.3 & 15.9
& 30.6 & 22.0
& 24.9 & 15.7
& 40.6 & 27.4 \\

-Dur
& 33.4 & 23.1
& 25.0 & 17.9
& 30.6 & 22.0
& 21.9 & 13.7
& 44.7 & 31.6 \\
\bottomrule
\end{tabular}
\vspace{-1em}
\end{table}

%% file: contents/5_conclusion.tex
\section{Conclusion}
We proposed \ours{} for text-based singing voice editing, combining melody-aware duration-ratio prediction with a flow-matching-based infilling decoder to generate melody-consistent edits while preserving total duration and non-edited regions. Experimental results show that \ours{} achieves state-of-the-art performance in both objective and subjective evaluations.
Future work includes developing melody-aware metrics and broader SVE settings. Samples are available at \url{https://cottonlove.github.io/MeloDISinger_demo/}.
\clearpage

%% file: contents/6_ack.tex
\section{Acknowledgments}
This work was supported by the National Research Foundation of Korea (NRF) grant funded by the Korea government (MSIT) (No. RS-
2023-00222383) and the Institute of Information \& communications
Technology Planning \& Evaluation (IITP) grant funded by the Korea government (MSIT) (No. RS-2019-II190075, Artificial Intelligence Graduate School Program (KAIST)).

%% file: mybib.bib
@article{sundberg2018singing,
  title={The singing voice},
  author={Sundberg, Johan},
  journal={The Oxford handbook of voice perception},
  pages={117--142},
  year={2018},
  publisher={Oxford University Press Oxford, UK}
}

@article{fastspeech2,
  title={Fastspeech 2: Fast and high-quality end-to-end text to speech},
  author={Ren, Yi and Hu, Chenxu and Tan, Xu and Qin, Tao and Zhao, Sheng and Zhao, Zhou and Liu, Tie-Yan},
  journal={arXiv preprint arXiv:2006.04558},
  year={2020}
}

@inproceedings{editsinger,
  title={EditSinger: Zero-Shot Text-Based Singing Voice Editing System with Diverse Prosody Modeling.},
  author={Zhang, Lichao and Zhao, Zhou and Ren, Yi and Deng, Liqun},
  booktitle={IJCAI},
  pages={4503--4509},
  year={2022}
}

@article{vevo2,
  title        = {Vevo2: A Unified and Controllable Framework for Speech and Singing Voice Generation},
  author       = {Zhang, Xueyao and Zhang, Junan and Wang, Yuancheng and Wang, Chaoren and Chen, Yuanzhe and Jia, Dongya and Chen, Zhuo and Wu, Zhizheng},
  journal      = {{IEEE} {ACM} Trans. Audio Speech Lang. Process.},
  year         = {2026}
}

@article{yingmusic,
  title={YingMusic-Singer: Zero-shot Singing Voice Synthesis and Editing with Annotation-free Melody Guidance},
  author={Zheng, Junjie and Hao, Chunbo and Ma, Guobin and Zhang, Xiaoyu and Chen, Gongyu and Ding, Chaofan and Chen, Zihao and Xie, Lei},
  journal={arXiv preprint arXiv:2512.04779},
  year={2025}
}

@misc{alexos2024attentionstitchattentionsolvesspeech,
      title={AttentionStitch: How Attention Solves the Speech Editing Problem}, 
      author={Antonios Alexos and Pierre Baldi},
      year={2024},
      eprint={2403.04804},
      archivePrefix={arXiv},
      primaryClass={eess.AS},
      url={https://arxiv.org/abs/2403.04804}, 
}

@inproceedings{peng2024voicecraft,
  title={Voicecraft: Zero-shot speech editing and text-to-speech in the wild},
  author={Peng, Puyuan and Huang, Po-Yao and Li, Shang-Wen and Mohamed, Abdelrahman and Harwath, David},
  booktitle={Proceedings of the 62nd Annual Meeting of the Association for Computational Linguistics (Volume 1: Long Papers)},
  pages={12442--12462},
  year={2024}
}

@article{lei2024songcreator,
  title={Songcreator: Lyrics-based universal song generation},
  author={Lei, Shun and Zhou, Yixuan and Tang, Boshi and Lam, Max WY and Liu, Hangyu and Wu, Jingcheng and Kang, Shiyin and Wu, Zhiyong and Meng, Helen and others},
  journal={Advances in Neural Information Processing Systems},
  volume={37},
  pages={80107--80140},
  year={2024}
}

@inproceedings{gtsinger, series={NeurIPS 2024},
   title={GTSinger: A Global Multi-Technique Singing Corpus with Realistic Music Scores for All Singing Tasks},
   url={http://dx.doi.org/10.52202/079017-0034},
   DOI={10.52202/079017-0034},
   booktitle={Advances in Neural Information Processing Systems 37},
   publisher={Neural Information Processing Systems Foundation, Inc. (NeurIPS)},
   author={Chen, Yuxin and Cheng, Xinyu and Guo, Wenxiang and He, Jinzheng and Hong, Zhiqing and Jiang, Ziyue and Li, Ruiqi and Lu, Jingyu and Pan, Changhao and Wang, Chuxin and Wang, Jialei and Xu, Wenhao and Yang, Chen and Zhang, LiChao and Zhang, Yu and Zhao, Zhou and Zhou, Jiecheng and Zhu, Zhiyuan},
   year={2024},
   pages={1117–1140},
   collection={NeurIPS 2024} }

@misc{hifisinger,
      title={HiFiSinger: Towards High-Fidelity Neural Singing Voice Synthesis}, 
      author={Jiawei Chen and Xu Tan and Jian Luan and Tao Qin and Tie-Yan Liu},
      year={2020},
      eprint={2009.01776},
      archivePrefix={arXiv},
      primaryClass={eess.AS},
      url={https://arxiv.org/abs/2009.01776}, 
}

@inproceedings{whisper,
  title={Robust speech recognition via large-scale weak supervision},
  author={Radford, Alec and Kim, Jong Wook and Xu, Tao and Brockman, Greg and McLeavey, Christine and Sutskever, Ilya},
  booktitle={International conference on machine learning},
  pages={28492--28518},
  year={2023},
  organization={PMLR}
}

@inproceedings{fpc_1,
  title={The singing voice conversion challenge 2023},
  author={Huang, Wen-Chin and Violeta, Lester Phillip and Liu, Songxiang and Shi, Jiatong and Toda, Tomoki},
  booktitle={2023 IEEE Automatic Speech Recognition and Understanding Workshop (ASRU)},
  pages={1--8},
  year={2023},
  organization={IEEE}
}

@inproceedings{visinger,
  title={Visinger: Variational inference with adversarial learning for end-to-end singing voice synthesis},
  author={Zhang, Yongmao and Cong, Jian and Xue, Heyang and Xie, Lei and Zhu, Pengcheng and Bi, Mengxiao},
  booktitle={ICASSP 2022-2022 IEEE International Conference on Acoustics, Speech and Signal Processing (ICASSP)},
  pages={7237--7241},
  year={2022},
  organization={IEEE}
}

@article{xiaoicesing,
  title={Xiaoicesing: A high-quality and integrated singing voice synthesis system},
  author={Lu, Peiling and Wu, Jie and Luan, Jian and Tan, Xu and Zhou, Li},
  journal={arXiv preprint arXiv:2006.06261},
  year={2020}
}

@inproceedings{GA_1,
   title={Efficiently Trainable Text-to-Speech System Based on Deep Convolutional Networks with Guided Attention},
   url={http://dx.doi.org/10.1109/ICASSP.2018.8461829},
   DOI={10.1109/icassp.2018.8461829},
   booktitle={2018 IEEE International Conference on Acoustics, Speech and Signal Processing (ICASSP)},
   publisher={IEEE},
   author={Tachibana, Hideyuki and Uenoyama, Katsuya and Aihara, Shunsuke},
   year={2018},
   month=apr, pages={4784–4788} }

@inproceedings{GA_2,
  title={Deepsinger: Singing voice synthesis with data mined from the web},
  author={Ren, Yi and Tan, Xu and Qin, Tao and Luan, Jian and Zhao, Zhou and Liu, Tie-Yan},
  booktitle={Proceedings of the 26th ACM SIGKDD International Conference on Knowledge Discovery \& Data Mining},
  pages={1979--1989},
  year={2020}
}

@article{voicebox,
  title={Voicebox: Text-guided multilingual universal speech generation at scale},
  author={Le, Matthew and Vyas, Apoorv and Shi, Bowen and Karrer, Brian and Sari, Leda and Moritz, Rashel and Williamson, Mary and Manohar, Vimal and Adi, Yossi and Mahadeokar, Jay and others},
  journal={Advances in neural information processing systems},
  volume={36},
  pages={14005--14034},
  year={2023}
}

@inproceedings{whisperx,
  title     = {{WhisperX: Time-Accurate Speech Transcription of Long-Form Audio}},
  author    = {Max Bain and Jaesung Huh and Tengda Han and Andrew Zisserman},
  year      = {2023},
  booktitle = {{Interspeech 2023}},
  pages     = {4489--4493},
  doi       = {10.21437/Interspeech.2023-78},
  issn      = {2958-1796},
}

@inproceedings{mfa,
  title={Montreal forced aligner: Trainable text-speech alignment using kaldi.},
  author={McAuliffe, Michael and Socolof, Michaela and Mihuc, Sarah and Wagner, Michael and Sonderegger, Morgan},
  booktitle={Interspeech},
  volume={2017},
  pages={498--502},
  year={2017}
}

@inproceedings{diffsinger,
  title={Diffsinger: Singing voice synthesis via shallow diffusion mechanism},
  author={Liu, Jinglin and Li, Chengxi and Ren, Yi and Chen, Feiyang and Zhao, Zhou},
  booktitle={Proceedings of the AAAI conference on artificial intelligence},
  volume={36},
  number={10},
  pages={11020--11028},
  year={2022}
}

@article{wavenet,
  title={Wavenet: A generative model for raw audio},
  author={Van Den Oord, Aaron and Dieleman, Sander and Zen, Heiga and Simonyan, Karen and Vinyals, Oriol and Graves, Alex and Kalchbrenner, Nal and Senior, Andrew and Kavukcuoglu, Koray and others},
  journal={arXiv preprint arXiv:1609.03499},
  number={1},
  year={2016}
}

@inproceedings{wang2022singingtacotron,
author = {Wang, Tao and Fu, Ruibo and Yi, Jiangyan and Wen, Zhengqi and Tao, Jianhua},
title = {Singing-Tacotron: Global Duration Control Attention and Dynamic Filter for End-to-end Singing Voice Synthesis},
year = {2022},
isbn = {9781450394963},
publisher = {Association for Computing Machinery},
address = {New York, NY, USA},
url = {https://doi.org/10.1145/3552466.3556534},
doi = {10.1145/3552466.3556534},
booktitle = {Proceedings of the 1st International Workshop on Deepfake Detection for Audio Multimedia},
pages = {53–59},
numpages = {7},
location = {Lisboa, Portugal},
series = {DDAM '22}
}

@INPROCEEDINGS{tacotron,
  author={Shen, Jonathan and Pang, Ruoming and Weiss, Ron J. and Schuster, Mike and Jaitly, Navdeep and Yang, Zongheng and Chen, Zhifeng and Zhang, Yu and Wang, Yuxuan and Skerrv-Ryan, Rj and Saurous, Rif A. and Agiomvrgiannakis, Yannis and Wu, Yonghui},
  booktitle={2018 IEEE International Conference on Acoustics, Speech and Signal Processing (ICASSP)}, 
  title={Natural TTS Synthesis by Conditioning Wavenet on MEL Spectrogram Predictions}, 
  year={2018},
  volume={},
  number={},
  pages={4779-4783},
  doi={10.1109/ICASSP.2018.8461368}}

@article{Eskimez2024E2TE,
  title={E2 TTS: Embarrassingly Easy Fully Non-Autoregressive Zero-Shot TTS},
  author={Sefik Emre Eskimez and Xiaofei Wang and Manthan Thakker and Canrun Li and Chung-Hsien Tsai and Zhen Xiao and Hemin Yang and Zirun Zhu and Min Tang and Xu Tan and Yanqing Liu and Sheng Zhao and Naoyuki Kanda},
  journal={2024 IEEE Spoken Language Technology Workshop (SLT)},
  year={2024},
  pages={682-689},
  url={https://api.semanticscholar.org/CorpusID:270738197}
}

@article{chen-etal-2024-f5tts,
      title={F5-TTS: A Fairytaler that Fakes Fluent and Faithful Speech with Flow Matching}, 
      author={Yushen Chen and Zhikang Niu and Ziyang Ma and Keqi Deng and Chunhui Wang and Jian Zhao and Kai Yu and Xie Chen},
      journal={arXiv preprint arXiv:2410.06885},
      year={2024},
}

@inproceedings{Zhang_2024,
   title={TCSinger: Zero-Shot Singing Voice Synthesis with Style Transfer and Multi-Level Style Control},
   url={http://dx.doi.org/10.18653/v1/2024.emnlp-main.117},
   DOI={10.18653/v1/2024.emnlp-main.117},
   booktitle={Proceedings of the 2024 Conference on Empirical Methods in Natural Language Processing},
   publisher={Association for Computational Linguistics},
   author={Zhang, Yu and Jiang, Ziyue and Li, Ruiqi and Pan, Changhao and He, Jinzheng and Huang, Rongjie and Wang, Chuxin and Zhao, Zhou},
   year={2024},
   pages={1960–1975} }

@article{wu2024laugh,
  title={Laugh Now Cry Later: Controlling Time-Varying Emotional States of Flow-Matching-Based Zero-Shot Text-to-Speech},
  author={Wu, Haibin and Wang, Xiaofei and Eskimez, Sefik Emre and Thakker, Manthan and Tompkins, Daniel and Tsai, Chung-Hsien and Li, Canrun and Xiao, Zhen and Zhao, Sheng and Li, Jinyu and others},
  journal={IEEE Spoken Language Technology Workshop (SLT)},
  year={2024}
}

@article{lipman2022flow,
  title={Flow matching for generative modeling},
  author={Lipman, Yaron and Chen, Ricky TQ and Ben-Hamu, Heli and Nickel, Maximilian and Le, Matt},
  journal={arXiv preprint arXiv:2210.02747},
  year={2022}
}
